# Learning warps object representations in the ventral temporal cortex

Alex Clarke[1,2], Philip J. Pell[1], Charan Ranganath[2], Lorraine K. Tyler[1]
[1]Centre for Speech, Language and the Brain, Department of Psychology, University of Cambridge, UK. CB2 3EB.
[2]Center for Neuroscience, University of California Davis, Davis CA, 95618.

## Abstract

The human ventral temporal cortex (VTC) plays a critical role in object recognition. Although it is well established that visual experience shapes VTC object representations, the impact of semantic and contextual learning is unclear. In this study, we tracked changes in representations of novel visual objects that emerged after learning meaningful information about each object. Over multiple training sessions, participants learned to associate semantic features (e.g. "made of wood", "floats") and spatial contextual associations (e.g. "found in gardens") with novel objects. Functional magnetic resonance imaging was used to examine VTC activity for objects before and after learning. Multivariate pattern similarity analyses revealed that, after learning, VTC activity patterns carried information about the learned contextual associations of the objects, such that objects with contextual associations exhibited higher pattern similarity after learning. Further, these learning-induced increases in pattern information about contextual associations were correlated with reductions in pattern information about the object's visual features. In a second experiment, we validated that these contextual effects translated to real-life objects. Our findings demonstrate that visual object representations in VTC are shaped by the knowledge we have about objects, and show that object representations can flexibly adapt as a consequence of learning with the changes related to the specific kind of newly acquired information.

# Introduction

The ventral temporal cortex (VTC) is crucial for object recognition (Clarke & Tyler, 2014; Grill-Spector et al., 1998; Kravitz, Saleem, Baker, Ungerleider, & Mishkin, 2013; Martin, 2007; Ungerleider & Mishkin, 1982) and many studies have demonstrated that visual experience can shape object representations in human VTC (see Kourtzi & Connor, 2011; Op de Beeck & Baker, 2010). In daily life, objects are not only processed according to their visual appearance, but also according to their meaning. Little is known, however, about how visual object representations change as they transition from being meaningless to meaningful.

In order to understand how learning about meaning changes object representations, it is important to distinguish between different dimensions of meaning that could be learned. For instance, one can learn about intrinsic attributes (e.g. *has ears, made of metal, floats*) that determine the function or significance of an object, or about spatial contextual associations (e.g. *found in the zoo*) that enable objects to be situated in the world. Previous studies, in which participants learned semantic features for meaningless objects have shown learning-related increases in brain responses (Bellebaum et al., 2013; James & Gauthier, 2003, 2004; Moore, Cohen, & Ranganath, 2006; Skipper, Ross, & Olson, 2011; Vuilleumier, Henson, Driver, & Dolan, 2002; Weisberg, van Turennout, & Martin, 2007). These studies show that learning object meaning can change how objects are represented in VTC, but they do not address the central issue of how specific aspects of meaning drive changes in the neural representation of objects.

The aim of the present study was to test if there is a relationship between the specific type of information people learned and how object representations changed. We used functional magnetic resonance imaging (fMRI) to examine how representations of pre-experimentally novel objects are modified as they become meaningful. We examined the effect of learning two aspects of meaning – an object's semantic category and its contextual association, both linked to regions of the VTC such as the posterior fusiform gyrus (Chao, Haxby, & Martin, 1999; Clarke & Tyler, 2014; Cox & Savoy, 2003; Huth, Nishimoto, Vu, & Gallant, 2012), parahipocampal cortex (PHC), and extending to the retrosplenial cortex (RSC; Aminoff, Gronau, & Bar, 2007; Bar & Aminoff, 2003; Bar, Aminoff, & Schacter, 2008; Stansbury, Naselaris, & Gallant, 2013). Further, we tracked how learning about different aspects of meaning impacts on visual shape-based representations.

Participants learned a name and four semantic features for each of 12 novel objects, that created a semantic category structure (three different categories based on overlap of features), while objects either did or did not have a contextual feature. We used multivariate representational similarity analysis (RSA; Kriegeskorte, Mur, & Bandettini, 2008; Nili et al., 2014) to determine how learning these two aspects of meaning - semantic category and contextual associations- effects neural similarity spaces in VTC. We predict that if learning meaningful information drives changes in neural similarity spaces, then objects from the same semantic category, or those associated with a context, will show more similar activation patterns in VTC than objects not sharing that property. Within the VTC, we predict semantic category effects to be most prominent in the fusiform, while contextual association effects are predicted to be more widespread including the RSC, PHC, lingual and fusiform. Finally, the visual shape similarity of the objects is predicted to relate to activation patterns in early visual cortex possibly extending into the VTC, and allows us to test for a relationship between visual form information and the newly learned semantic information.

To assess the generalizability of our results, we ran a second experiment to examine how semantic category and contextual associations influence VTC activity patterns during processing of real-world objects. Based on previous studies we would expect category and contextual effects for real objects, and their inclusion here not only demonstrates the generalisability of any learning effects, but importantly allows for comparisons of the distribution of effects for real and novel objects which can help establish whether effects of our learning paradigm lead to the kinds of representational changes that mirror the long-term learned representations that are present for real objects.

# Methods

## Experiment 1: Novel objects

### Overview
An overview of the experimental sessions can be seen in Table 1. The experiment consisted of two identical fMRI sessions (mean time between scans was 26 days; range 21-28 days) and four behavioural learning/testing sessions during which the semantic feature information was learned for 12 novel objects. During the fMRI sessions, participants performed a simple visual task that could be performed on the objects both before and after learning, and all behavioural sessions were conducted between the two fMRI sessions. The behavioural sessions were completed within one week, with behavioural session 3 occurring between 48 and 72 hours prior to the second fMRI session, and behavioural session 4 occurring just prior to the second scanning session.

### Participants
Twelve healthy participants (6 males, 6 females) completed all MRI and behavioural sessions. All had normal or corrected to normal vision and were right handed. The average age was 20.7 years (range 19-23 years). All participants provided informed consent, and the study was approved by the Cambridge Research Ethics Committee. One participant was excluded due to excessive head movement during both scanning sessions, and was excluded from all analyses (leaving eleven participants).

### Stimuli
A total of 24 non-real, novel objects were used in the study - 12 were assigned to the learning condition and 12 to the exposure condition. The novel objects were "fribbles" (downloaded from http://wiki.cnbc.cmu.edu/Novel_Objects), and have a main body and four appendages. Eight fribbles were selected from three different visual "species", where a species is defined by a common main body. Within each species, two fribbles were selected from each of four "families", where members of a family share the main body and the appendages have a variable degree of overlap, and fribbles from the same species but different families share a main body but have no appendages in common. This structure provides a variable degree of visual similarity across the novel objects based on the main body and appendages (visual features). Half of the fribbles were used for the learning condition and half for the exposure condition. In each condition, and for all species, the amount of visual feature overlap was matched to ensure that effects were due to the semantic, rather than the visual, properties of the objects.

All fribbles were displayed as grey-scale images in the centre of a white background. Pretesting was used to ensure that the 24 fribbles did not have a strong resemblance to familiar object categories (animals, plant life, tools, vehicles, living, nonliving). Eight items from the exposure set were inverted based on pretests so they no longer showed a resemblance to familiar object categories.

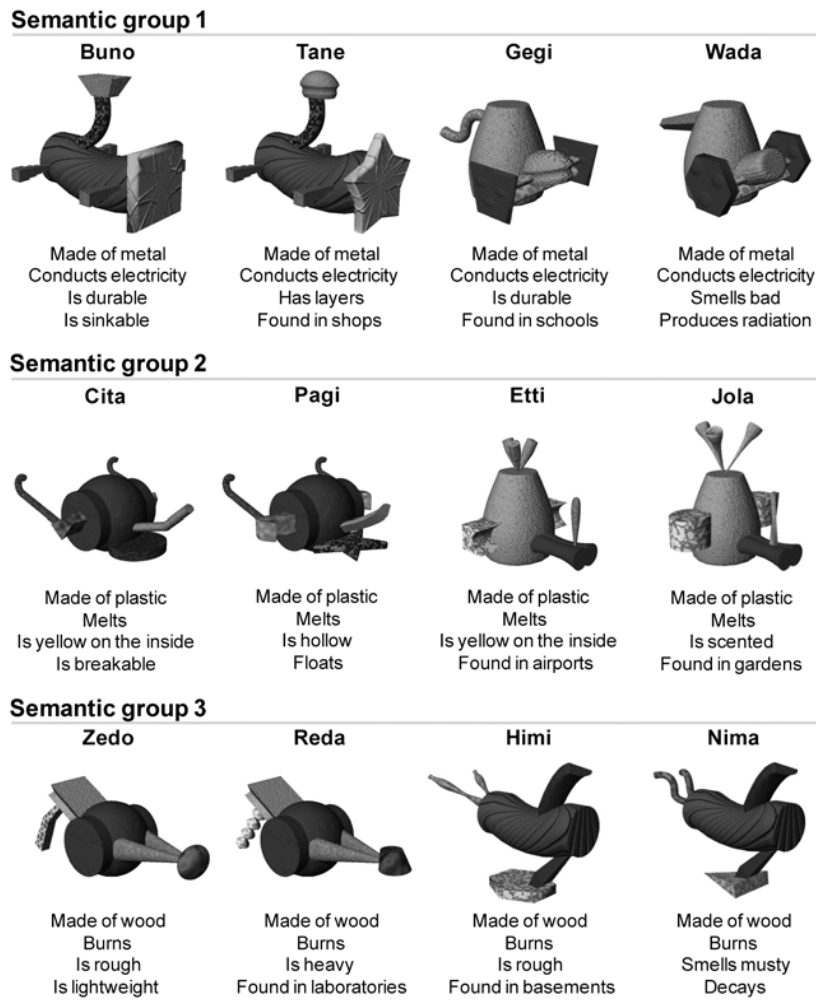

*Figure 1. Objects in the learning condition, showing their names and semantic features. Each row shows a different semantic group, where each member shares 2 or 3 features. Each semantic group is composed of objects from two different visual groups.*

The 12 fribbles from the learning condition were each assigned a name and four semantic features (e.g. *made of metal*; Figure 1) forming three semantic groups based on semantic feature overlap. Each semantic group contained four fribbles that were drawn from two different visual species. All items within a semantic group had two semantic features in common that were not present for items in other semantic groups (the shared features of that group i.e. Made of metal & Conducts electricity for the four fribbles on the top row of Figure 1). These shared features provide the basis for semantic similarity within each group and the basis for category organisation. Further, one semantic feature was shared between two items in each group (less shared features; e.g. Is durable is present in two fribbles on the top row of Figure 1) and all other semantic features were unique, with a total of 27 features. Twenty six of the semantic features were selected from the McRae et al., (2005) feature production norms, and one feature was chosen by the experimenters.

All semantic features were non-visual (i.e., not describing visual object form) and synonyms were avoided to ensure features were easily discriminable. Features that signify animacy (e.g., eats) or specify the function of familiar items (e.g., used for holding liquids) were also avoided to increase the plausibility of the features for novel objects. The two shared semantic features were a material (e.g., made of wood) and a well known property of that material (e.g., burns). Two of the unique semantic features in each semantic group were contextual features (e.g. Found in...). The remaining semantic features were sensory (e.g., smells bad), hidden-visual (e.g., yellow on the inside) or general object properties (e.g. has layers, is lightweight). This selection of the fribbles and assignment of semantic features allows us to test for high-level visual, semantic feature and contextual similarities in the brain due to learning.

**Procedure**

*Behavioural sessions*
Participants learned to associate semantic information for 12 novel objects over two learning/testing sessions on separate days (behavioural sessions 1 to 2) and two testing session (behavioural sessions 3 to 4). While learning was only given in the first two sessions, testing was conducted in all behavioural sessions to track learning rates over time.

*Learning sessions*
The 12 objects were learned in subsets of 4 items. Initially participants viewed information slides for each object in the subset. Slides contained an image of the object along with text containing its name and semantic features (Figure 2a). Participants were instructed they would be tested on their knowledge of the text and were asked to read each slide carefully. After viewing the information slides, participants answered a series of questions about the objects and their features (Figure 2b). Questions were presented in three phases; Phase 1 - associating objects with the shared features. Phase 2 - associating objects with less shared and unique features. Phase 3 - associating names with the objects/semantic features. Questions followed the general form of presenting a target and two response options. The nature of the targets and response options varied across trials and phases but could be an object image, name, single feature or pair of features. The response options remained on screen until a decision was made, and feedback was given on each trial to reinforce learning. Information slides for each object were intermixed within the question trials. Each phase was repeated if accuracy was below 80%. In total there were 252 question trials in behavioural session 1 and 216 question trials in behavioural session 2, where only phases 2 and 3 were given.

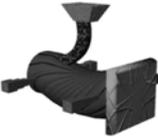

*Figure 2. Semantic learning and testing. a) Example information slide presented to participants in order to learn semantic information. b) Example question trial from phase 2 of learning. c) Example question from the forced-choice recognition task in the testing phase*

### Testing

Two methods were used to measure learning performance: a forced-choice recognition task and a feature-recall task. Forced-choice recognition tests were administered after learning in behavioural sessions 1 and 2, and again during behavioural session 3 (where no learning took place). On each trial, four semantic features belonging to the same object were presented with three options (Figure 2c), that were either object images or names. The incorrect options were either from the same or different semantic groups. There were a total of 48 trials in the testing sessions. Semantic features and response options remained on screen until a response was made, and no feedback was given. The lowest accuracy across all three testing sessions was 79% (Table 2), with all participants scoring higher than 85% in the final testing session and 5/11 participants scoring 100%.

Feature-recall tests were administered at the start of behavioural session 2, 3 and 4. Participants were given two sheets of paper containing the images of the 12 learned, novel objects, above five empty lines labelled: "Name", "Feature 1", "Feature 2", "Feature 3", and "Feature 4". Participants were asked to fill in as much information as possible within a 10 minute time limit. The ordering and location of each object on the answer sheets was randomised across the three tests, and was not ordered by semantic or visual groupings. The lowest accuracy in the final testing session (behavioural session 4) was 67%, while 7/11 participants scored 100% (Table 2). Overall, the testing scores indicate that the vast majority of participants learned the associations between the semantic features and the object images to a high level of success (9/11 scored at least 98% in the forced-choice testing, and 8/11 scored at least 96% in the free-recall).

### *Exposure condition*

An additional 12 novel objects were included to act as a control condition for any influence of visual familiarity and to test for overall effects of learning meaning. These objects were not associated with semantic features, but participants were exposed to them using a one-back visual matching task that was performed during behavioural sessions 2 and 3. On each trial an object appeared for 600 ms and was followed by a 1 s central fixation cross. Participants were instructed to press the right button if the object on screen did not match the previous item or the left button if it did match the previous item. A 12-trial practice of the one-back task was followed by two 96-trial blocks. Within each block there were 8 repetitions of each item and each item was twice a 'match' target; giving a total of 24 match-trials in each block. Participants were given feedback on their performance at the end of each block.

### fMRI sessions

### *Procedure*

Both pre and post-learning fMRI sessions used the same task and procedure. Participants performed a visual anomaly detection task where they had to detect when one of the objects' visual features was 'bleached out'. The task ensured participants paid close attention to the images, could perform the task both prior to and post-learning, and could be equally performed with objects from the learning and exposure sets. Two modified versions of each of the 24 objects was created for the anomaly detection task by increasing the brightness and contrast of one of the appendage features by 60% using Adobe Photoshop CS2. A different feature was modified in each version. One version was used in the first scanning session and the other in the second scanning session.

Each trial consisted of a centrally presented black fixation cross on a white background for 300 ms, followed by a picture lasting 700ms, then a blank white background between 2 s and 7 s. The participants' task was to press one button if the object on screen was an unmodified image or another button if the object was a modified image (where a feature was bleached out; see Fig M4 centre panel below). There were 15 repetitions of each image; 12 unmodified versions and three modified versions. Objects were presented in 15 blocks each containing a single presentation of all 24 object with four or five modified images in each block. The order of items within a block was random, and different for every block. The position of the four/five modified items in each block was random with the constraint that no more than two could occur in succession. The presentation and timing of stimuli was controlled using Eprime version 1.1 (Psychology Software Tools, Pittsburgh, PA).

## *Scanning*

All scanning took place at the MRC Cognition and Brain Sciences Unit, Cambridge, in a Siemens 3-T Tim Trio MRI scanner (Siemens Medical Solutions, Camberley, UK). Two functional scans were collected in each scanning session using gradient-echo echoplanar imaging (EPI) sequences collecting 32 slices in descending order of 3 mm thickness and between slice gap of 0.75 mm and a resolution of 3 x 3 mm. The field-of-view was 192 x 192 mm, matrix size 64 x 64 with a TE of 30 ms, TR of 2 s and a flip angle of 78⁰. Each functional scan lasted approximately 12 minutes with a short break half-way through. Prior to functional scanning, a high-resolution structural MRI image was collected using an MPRAGE sequence with 1 mm isotropic resolution.

## *fMRI preprocessing*

Data from the two scanning sessions were preprocessed independently. Functional images were slice-time corrected, spatially realigned and smoothed using a 4 mm Gaussian kernel in SPM8 (Wellcome Institute of Cognitive Neurology, London, UK). These unnormalised images were analysed for each participant with the general linear model to creating a single beta image for each object based on the 12 repetitions of the unmodified images. In addition to the 24 novel object predictors in the GLM, predictors were included to capture effects associated with the modified objects, slow trends using 11 regressors for the first block of functional scans and 12 for the second block based on the basis functions of a discrete cosine transform (minimum frequency = 1/128 Hz), six head motion regressors for each session along with their first derivatives, and a global mean predictor for each scanning session. We also included a separate regressor for each fast-motion event, where a fast motion event was defined as motion greater than 0.7 mm/TR and detected using ArtRepair software (Mazaika, Hoeft, Glover, & Reiss, 2009). The resulting 24 beta images for the unmodified objects were converted to t-images prior to the RSA analyses.

## *RSA analysis*

We first used a searchlight mapping approach (Kriegeskorte, Goebel, & Bandettini, 2006) to test for either consistent effects over sessions, or learning-induced changes to how the novel objects were represented in the pre- and post-learning fMRI sessions. Searchlight analyses were followed by analyses of anatomically defined ROIs that have been previously linked to semantic category and contextual effects, and was performed as activation patterns may encode information at the spatial scale of anatomical regions which searchlight analyses may be less sensitive to due to their smaller spatial neighbourhood (Etzel, Zacks, & Braver, 2013).

Candidate model RDMs

We tested a visual model RDM and three other models capturing different semantic distinctions (Figure 3). The *visual features* model captures the amount of visual-feature overlap between pairs of objects, where a feature can be the main body or an appendage. This is a high-level model of visual similarity that captures visual shape information, rather than simply low-level visual information. Objects from the same fribble species will always share at least one feature (the main body), while objects from different fribble species always share no features. The *meaning* model tests for overall effects of learning meaning in contrast to the objects in the exposure condition. Here, objects that were learned about are predicted to cluster together to a greater degree that objects from the exposure condition, which have no associated meaning. The *semantic category* model captures the three semantic categories the objects belonged to, where members of the same semantic category share either 2 or 3 features, and no features were shared across semantic categories. The *contextual* model tests where activation patterns for the learned objects form two clusters according to whether they were associated with a specific context or have no contextual association (the correlation between the semantic category and contextual association RDMs was r = 0.19).

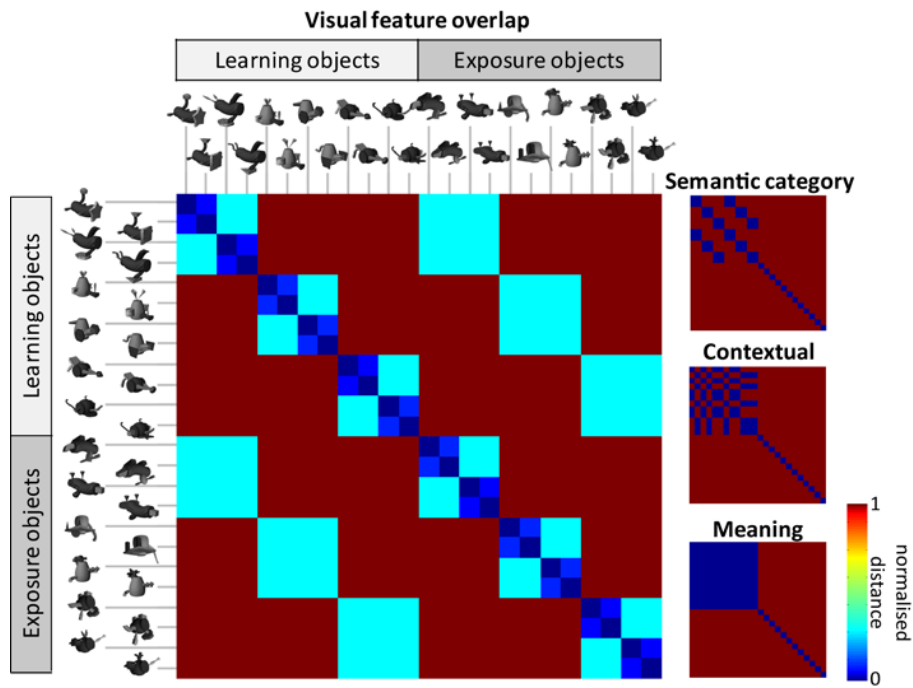

*Figure 3. Candidate model RDMs tested using representational similarity analysis.*

Searchlight

Each candidate model RDM was tested against the observed activation patterns using a searchlight similarity analysis implemented in the RSA toolbox (Nili et al., 2014) and custom matlab functions, before we tested for significant conjunctions or changes over sessions. At each voxel, object activation values from grey matter voxels within a spherical searchlight (radius 9 mm) were extracted to calculate distances between all items (using 1 - Pearson correlation) creating an object dissimilarity matrix based on that searchlight. This fMRI RDM was correlated with each candidate model RDM (using Spearman's rank correlation) and the resulting similarity values were Fisher transformed and mapped back to the voxel at the centre of the searchlight. Similarity maps for each model RDM and each participant were normalised to the MNI template space and spatially smoothed using a 6 mm FWHM Gaussian kernel. The similarity maps for each participant were then entered into a group-level random effects (RFX) analyses using SPM8. We used a paired samples t test design testing for first, common RSA effects across both sessions using a conjunction analysis (the conjunction null hypothesis) for positive effects, and second, increased RSA effects in the post-learning session compared to the pre-learning session. Given our *a priori* expectation that effects will be in the ventral processing stream, these analyses were constrained to a ventral stream mask (including the RSC) produced by combining bilateral regions from the Harvard-Oxford brain atlas (occipital pole, intracalcarine cortex, cuneal cortex, lateral occipital inferior, fusiform occipital, fusiform temporooccipital, lingual gyrus, parahippocampal posterior) and the RSC (BA29 & 30). Results are reported using a voxelwise threshold of $p < 0.005$ and a cluster extent of $p < 0.05$ (FWE) correcting for multiple comparisons.

ROI analysis

An anatomical ROI analysis was performed to test for effects of the candidate model RDMs in targeted VTC regions that have been previously implicated in semantic category and contextual processing. This was done to test for distributed regional effects that may only be present in activation patterns at the spatial scale of regions, rather than the smaller searchlights. This would be the case if information was distributed across voxels at a spatial scale larger than the searchlights. A number of ROIs were specified to cover the VTC, specifically the fusiform (fusiform temporooccipital), lingual (lingual gyrus) and parahippocampal (parahippocampal posterior). ROIs were defined from the Harvard-Oxford brain atlas (Desikan et al., 2006), and the retrosplenial cortex was also included and defined as BA29 and BA 30. For each ROI, object activation values were extracted for all voxels in the ROI to calculate distances between the 24 items (using 1 - Pearson correlation). The fMRI ROI RDMs were correlated with the candidate model RDM (using Spearman's rank correlation) and the resulting similarity values were Fisher transformed. RFX analyses were performed using one-tailed one-sampled t tests against zero when testing for significant positive effects of each model RDM, and paired-samples t-tests when testing for significant differences between fMRI sessions or model RDMs.

## Experiment 2: Real objects

### Participants

Sixteen participants took part in the study (6 male, 10 female). All were right-handed and were aged between 19 and 29 (mean 23 years). All participants had normal or corrected to normal vision, and gave informed consent. The study was approved by the Cambridge Research Ethics Committee.

## Stimuli

A total of 145 real objects were used, where 131 of these were from one of six object categories (34 animals, 15 fruit, 21 vegetables, 27 tools, 18 vehicles, 16 musical instruments) and 14 additional objects that did not adhere to a clear category (and were not included in our analyses). Isolated coloured objects were shown in the centre of a white background, and normalised to a maximum visual angle of 7.5°. All objects were chosen to depict concepts from an anglicised version of the McRae production norms (McRae et al., 2005; Taylor, Devereux, Acres, Randall, & Tyler, 2012) from which semantic feature information could be obtained to construct the meaningful model RDMs.

## Procedure

Participants performed an overt basic-level naming task. Each trial consisted of a fixation cross lasting 500 ms, before an object for 500 ms followed by a blank screen lasting between 3 and 11 seconds. All objects were repeated 6 times across 6 different blocks. The object presentation order for each block was randomised for each participant, although a constant category order was maintained ensuring an even distribution of object category across the block. The presentation and timing of stimuli were controlled with Eprime version 1 (Psychology Software Tools, Pittsburgh, PA, USA), and naming accuracy was recorded by the experimenter during acquisition.

## fMRI acquisition

Participants were scanned at the MRC Cognition and Brain Sciences Unit, Cambridge, in a Siemens 3-T Tim Trio MRI scanner (Siemens Medical Solutions, Camberley, UK). There were 3 functional scanning sessions using gradient-echo echoplanar imaging (EPI) sequences collecting 32 slices in descending order of 3 mm thickness and between slice gap of 0.75 mm and a resolution of 3 x 3 mm. The field-of-view was 192 x 192 mm, matrix size 64 x 64 with a TR of 2 seconds, TE of 30 ms and a flip angle of 78°. Each functional session lasted approximately 9-10 minutes, containing two object blocks. Prior to functional scanning, a high-resolution structural MRI image was collected using an MPRAGE sequence with 1 mm isotropic resolution.

## fMRI preprocessing

Functional images were slice-time corrected, spatially realigned and smoothed using a 4 mm Gaussian kernel in SPM8 (Wellcome Institute of Cognitive Neurology, London, UK). These unnormalised images were analysed for each participant with the general linear model to create a single beta image for each object based on the 6 repetitions. In addition to the 145 object predictors, predictors were included to capture slow trends using 18 regressors for each session based on the basis functions of a discrete cosine transform (minimum frequency = 1/128 Hz), six head motion regressors for each session and a global mean predictor for each scanning session. The resulting 145 beta images were converted to t-images prior to the RSA analyses. Only objects named correctly on all six repetitions (86%, SE = 1.53%) were included in further analyses.

## RSA analysis

An RSA analysis was performed for the real objects using both searchlight mapping and an anatomical ROI analysis. The procedures were identical to those used in experiment 1, with the exception that here the searchlight RFX analysis was conducted using a one-sampled t-test testing for positive effects of the models with no *a priori* voxel restrictions.

We tested for RSA effects of a contextual association RDM and a semantic category RDM (note that the high-level visual features and meaning models were not tested, as they can not be defined for our real objects in the same manner as they are defined for the novel objects). The contextual association model RDM was defined for real objects in the same manner as we defined it for novel objects. Using the McRae et al. (2005) production norm data to extract semantic features for our real objects, we can determine if an object has contextual associations or not. Features that indicated an object has strong contextual associations took the form; associated with X, found in/near/on X, lives in X, used at/on X, where X signified a specific location (e.g. deserts, school, Spain). Based on this criterion, 44 out of 131 objects were determined to have spatial contextual associations. The contextual association model RDM for real objects was constructed in the same way as for novel objects, where objects with contextual associations will form one cluster, and objects without contextual associations are form a second cluster. The semantic category RDM captures the category structure of the 131 objects across 6 superordinate categories (animals, fruit, vegetables, tools, vehicles, musical instruments) and indicates there is more within-category similarity than between-category similarity. The correlation between the semantic category and contextual association RDMs was r = 0.09.

# Results

## Experiment 1: Novel objects

The goal of this study was to determine whether learning-induced changes in object representations are governed by the semantic category and contextual association information people learn, and whether this impacts on visual form representations. The similarity relations between the objects, as embodied in the model RDMs, capture the visual and meaningful (learned) similarity spaces that can be tested against brain activation patterns to track changes in the information represented due to learning. Here, we used RSA to uncover statistical correspondences between the visual and meaningful (i.e. learned) similarity structures (Figure 3) and activation pattern similarities across the different objects. We first explored whether effects of our candidate model RDMs showed significant learning-induced changes, or consistent effects across sessions, in their relationship to activation patterns. To do this we directly compared the RSA searchlight maps from the pre- and post-learning scanning sessions, where the pre-learning session acts as a baseline for how the brain responds to meaningless objects. We then present an anatomical ROI analysis to look for more spatially distributed effects that extend beyond the size of our searchlights.

Using searchlight RSA, the visual features model was found to show consistent significant effects across both sessions in the occipital lobe (peak MNI coordinate: 21, -88, -1; Figure 4a), showing that searchlights within the visual cortex responded to the same images in a similar manner both before and after learning. Multidimensional scaling (MDS) further illustrated that the regional activation patterns clustered by visual form similarity, with objects from the same visual groups (shown by different colours) falling closer together in this 2-dimensional space than objects from different visual groups.

Testing for increased RSA effects after learning compared to before, we found significant effects relating to the contextual association model. Representational changes were seen in the right VTC with two foci – one on the fusiform gyrus (change in Spearman's rho = 0.08; peak MNI coordinate: 39, -34, -24) and one spanning the collateral sulcus (change in Spearman's rho = 0.04; peak MNI coordinate: 24, -52, -16; Figure 4b). MDS plots further showed how activation patterns in both these areas dissociate along the contextual association dimension of the stimuli. No effects were seen for the other meaning related models. These results show that activation patterns in the VTC have been warped through learning, whereby the same stimuli are represented in a different manner after learning in comparison to before, while object activation patterns in the early visual regions remained more stable in that they were significantly related to the visual feature information in both sessions, that were not significantly different to one another (Note that equivalent results are found by inspecting pre- and post- learning searchlight maps separately).

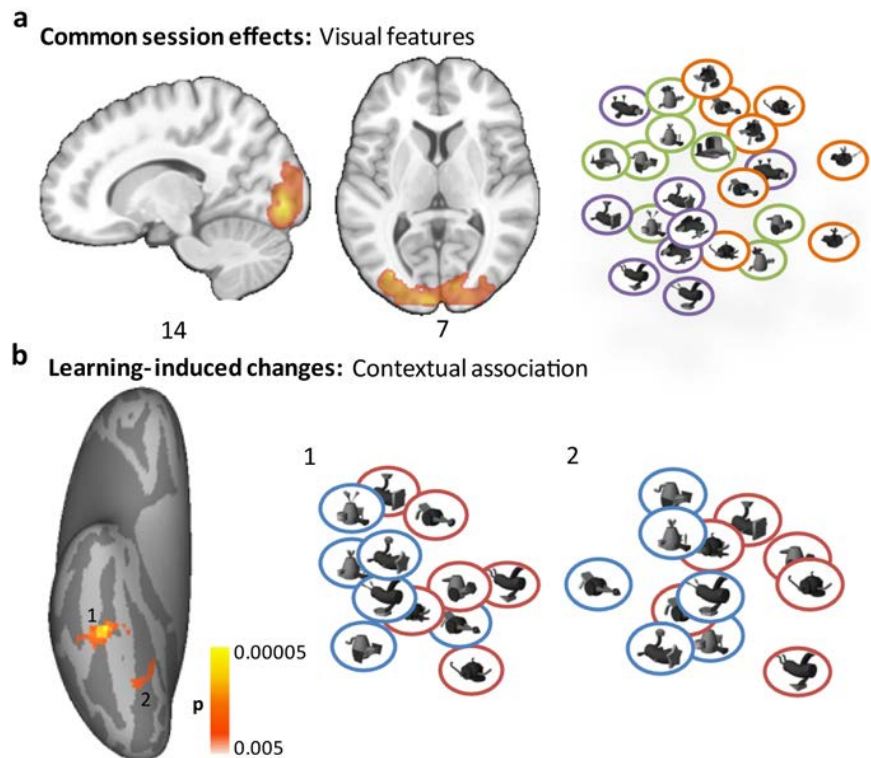

*Figure 4. RSA searchlight effects for novel objects. a) Searchlights showing common significant effects across both sessions for the visual features model. MDS plot derived from object pattern similarities averaged over sessions with each visual group shown in a different colour. b) Searchlights showing significant increases for the contextual associations model in the post-learning session compared to the pre-learning session. MDS plots derived from object pattern similarities in the post-learning session for the two foci, where objects with contextual associations shown in blue and without contextual associations in red (exposure objects not shown for clarity). Both images are voxelwise $p < 0.005$, cluster $p < 0.05$.*

While our searchlight analysis tested for effects in local activation patterns (i.e. searchlights), representational information could be present at the larger scale of anatomical regions which will not always be captured by smaller searchlights. This would be the case if information was distributed across voxels at a spatial scale larger than the searchlights. Therefore we also performed an analysis testing the candidate model RDMs within anatomically defined regions across the posterior ventral temporal lobe, including the RSC, that have been implicated in semantic category and contextual processing - specifically bilateral posterior fusiform, lingual, PHC and RSC (Figure 5).

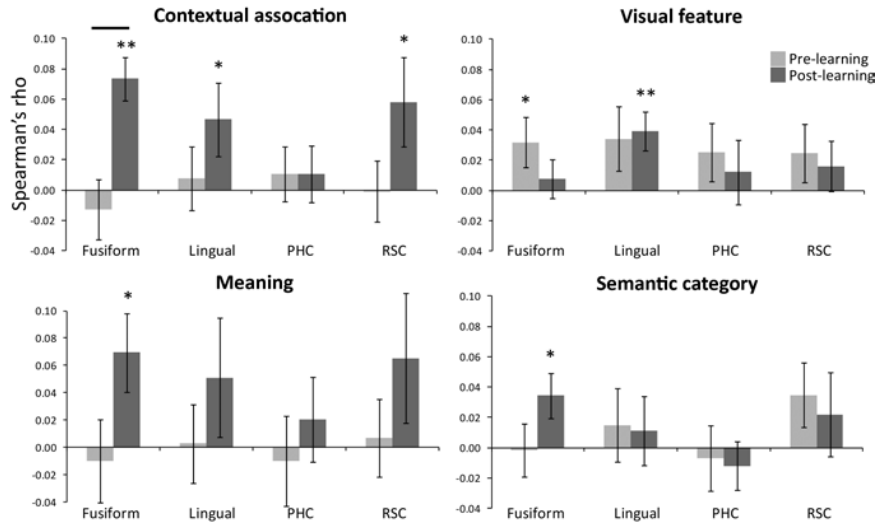

*Figure 5. RSA ROI analyses showing pre- and post-learning effects for the 4 model RDMs in the 4 anatomical ROIs. Pre-learning effects shown in light grey, post-learning effects in dark grey. Asterisks show significant effects of the models (\*\*p<0.01,\*p<0.05) and significant changes over sessions shown by a horizontal bar. PHC = parahippocampal cortex and RSC = retrosplenial cortex. Error bars show standard errors.*

Confirming the results from our searchlight analysis, we observed significant learning-induced increases in contextual association information in the fusiform in the post-learning session compared to the pre-learning session (t(10) = 3.37, p = 0.004). Further, the fusiform region showed significant effects of the contextual association model during the post-learning session (t(10) = 5.10, p = 0.0002), which were absent in the pre-learning session (t(10) = 0.6). We also observed marginally significant learning-induced increases for the contextual association model in the RSC (t(10) = 1.67, p = 0.064), which was significant for the post-learning session (t(10) = 1.96, p = 0.039) but not in the pre-learning session (t(10) = 0.05). Finally, the lingual region showed significant effects for the contextual association model in the post-learning session (t(10) = 1.92, p = 0.042), and not in the pre-learning session (t(10) = 0.4), although no significant change was seen between the two sessions (t(10) = 1.1).

Turning to the visual features model, a significant effect was seen in the fusiform for the pre-learning activation patterns (t(10) = 1.92, p = 0.042), that was reduced in the post-learning session and no longer significant (t(10) = 0.6), although cross session comparisons revealed no significant differences between the two sessions (t(10) = 0.85). The visual feature model also showed effects in the lingual region after learning (t(10) = 3.02, p = 0.006), which showed marginal effects in the pre-learning session (t(10) = 1.62, p = 0.068) that was not significantly different across the two scanning sessions (t(10) = 0.2) suggesting the lingual region reflected relatively stable object representations, similar to our searchlight results. Finally, we note that the other two meaning-related models – the meaning and semantic category RDMs, both showed significant effects in the post-learning session only (meaning: t(10) = 2.41, p = 0.018, semantic category: t(10) = 2.33, p = 0.021). These effects were absent in both the cross-session searchlight comparisons and in individual searchlight effects for the post-learning session (not shown) suggesting that information relating to general meaningfulness, and to the learned semantic categories the objects belong to, may be coded in more distributed patterns than the searchlight mapping was sensitive to.

The results from the anatomical ROI analyses confirm what was observed in the our searchlight analyses - learning about novel objects induced representational changes in VTC, most prominently in the fusiform gyrus, where patterns reflected the learned contextual associations of objects. Within the anatomical ROIs we also saw suggestions of a more general meaning-related effect conferred through the meaning and semantic category models. Moreover, after learning, visual similarity effects in the fusiform are reduced and learning-induced contextual similarities emerge. Testing the relationship between representational changes in the fusiform over the sessions showed a significantly greater change in RSA effects for the contextual model compared to both the visual feature model (t(10) = 2.2, p = 0.05) and the semantic category model (t(10) = 2.2, p = 0.05). Moreover, the reduction in the visual feature model effect was significantly correlated with the increased effect for the contextual association model (R-sq = 0.49, r = -0.7, p = 0.0085; Figure 6) indicating a representational shift in the kind of information that was represented in the fusiform as a consequence of learning.

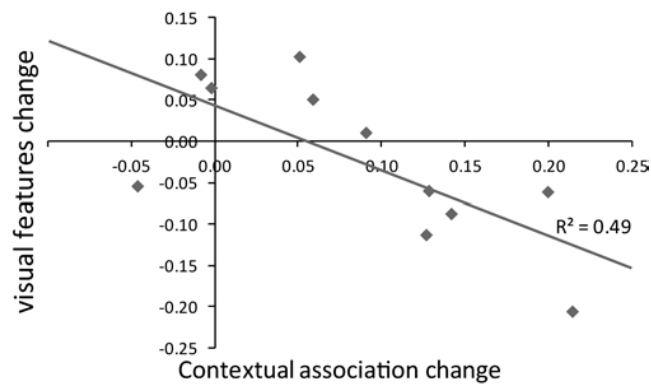

*Figure 6. Relationship of representational changes between pre- and post-learning sessions for visual feature and contextual association model RDMs.*

## Experiment 2: Real objects

In Experiment 1 we found that learning about contextual associations for pre-experimentally novel objects changed voxel pattern similarity information in VTC. However, in real life, contextual associations are learned by encountering objects in particular contexts, whereas for our novel objects associations were learned through reading text. Accordingly, in Experiment 2, we use RSA to test for effects of a contextual association model RDM for real objects in a previously published dataset (Clarke & Tyler, 2014). This not only allows us test the generalisability of our learning effects, but allows us to compare the distribution of effects for real and novel objects. Further we test for effects of a semantic category model, also known to show effects in the VTC, as a means of comparison, using both searchlight mapping and the anatomical ROIs within the VTC.

Searchlight RSA showed effects for the contextual association model primarily in bilateral posterior VTC including the fusiform, lingual, and parahippocampal cortices, and also including the calcarine and inferior occipital gyrus (peak MNI coordinate: -9, -91, 10; Figure 7a). The results echo those previously reported for the contextual associations of real world objects (e.g. Stansbury et al., 2013), validating the manner in which we define our contextual model, and partly overlapping with our effects for novel objects in the collateral sulcus. The semantic category effects were widespread throughout VTC (peak MNI coordinate: -27, -26, 59), similar to previously reported effects (e.g. Connolly et al., 2012; Huth et al., 2012) although we would also highlight that there are well known confounds between an object's semantic category and its visual properties, a factor that we controlled in our novel objects experiment.

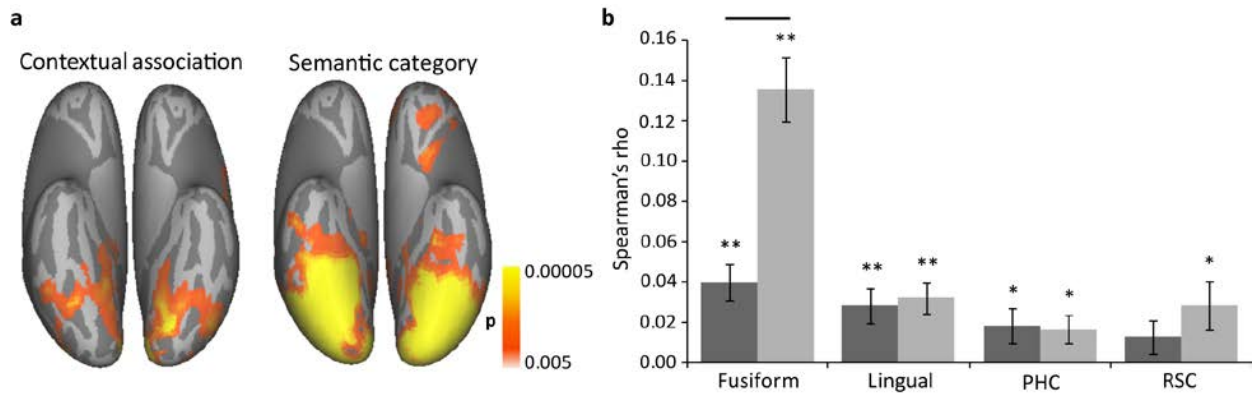

*Figure 7. RSA effects of the contextual association and semantic category models using real objects. a) Searchlight results, shown thresholded at voxelwise p < 0.005, cluster p < 0.05. b) RSA effects for the anatomical ROIs. Dark grey bars show effects for the contextual association model and light grey for the semantic category model. Asterisks show significant effects of the models (\*\*p<0.01,\*p<0.05), with significant changes over sessions shown by a horizontal bar. PHC = parahippocampal cortex, RSC = retrosplenial cortex, VTC = ventral temporal cortex. Error bars show standard errors.*

Within the anatomical ROIs, we found significant effects of the contextual association model in the fusiform (t(15) = 4.42, p = 0.0002), lingual (t(15) = 3.34, p = 0.0022) and PHC (t(15) = 2.13, p = 0.025) confirming our searchlight results are also present at this larger spatial scale (Figure 7b). The semantic category model showed significant effects in the fusiform (t(15) = 8.61, p < 0.0001), lingual (t(15) = 4.13, p = 0.0004), PHC (t(15) = 2.44, p = 0.014), and the RSC (t(15) = 2.34, p = 0.017). Further, the ROI similarity matrix was more similar to the semantic category model than the contextual association model in the fusiform ROI (t(15) = 5.10, p = 0.0001), showing the inverse relationship to that found for novel objects. Overall, both searchlight and ROI analyses showed activation patterns to real objects show a significant relationship to the contextual associations model with the strongest effects in the fusiform and lingual gyri, plus widespread semantic category effects across VTC that also peaked in the fusiform gyrus.

# Discussion

The goal of the present study was to determine how learning different kinds of meaningful information affects the neural representation of objects in VTC. In Experiment 1, we examined the effect of learning two aspects of meaning: the semantic features of an object, and whether each object had a specific contextual association (i.e., that an object is found in a particular setting). We found that learning about the presence of contextual associations induced systematic changes in the similarity structure of corresponding object representations in VTC. Learning led to the development of two clusters in the similarity space, one consisting of objects with a strong contextual association (i.e. a 'Found in...' feature) and another consisting of objects that did not have a strong contextual association. In Experiment 2, we demonstrated a distinction between images of real-world objects that had prominent contextual associations and objects that did not elicit strong contextual associations. Collectively, the results suggest that learned contextual associations exert powerful influences on the neural mechanisms of object processing. We elaborate on this and related issues below.

## Contextual association effects in VTC

Searchlight-based voxel pattern similarity analyses in Experiment 1 revealed that learning contextual associations affected object similarity spaces in VTC (Figure 4). Within VTC, there was a lateral peak in the fusiform gyrus and a more medial peak spanning the collateral sulcus where the similarity responses reflected the presence of learned contextual associations. The peak of the medial effect was posterior to the parahippocampal gyrus, peaking closely to reported scene selective responses (Aguirre, Zarahn, & D'Esposito, 1998). Anatomical ROI analyses converged with these results, showing contextual association effects for both novel and real objects in the fusiform, and changes for novel objects were also seen in RSC. Together, the findings demonstrate that learning about contextual associations causes significant representational changes in the VTC.

Large areas of the VTC have been shown to contain information about natural scene categories and their contextual associations (Stansbury et al., 2013; Walther, Caddigan, Fei-Fei, & Beck, 2009), in addition to meaningful object information more generally (Huth et al., 2012; Mahon, Anzellotti, Schwarzbach, Zampini, & Caramazza, 2009; Tyler et al., 2013; Vuilleumier et al., 2002). The emergence of learned information suggests that representations in the VTC are malleable through short term learning, with this area strongly implemented in previous learning studies (Gauthier, Tarr, Aanderson, Skudlarski, & Gore, 1999; James & Gauthier, 2004; Kourtzi, Betts, Sarkheil, & Welchman, 2005; Moore et al., 2006; Op de Beeck, Baker, DiCarlo, & Kanwisher, 2006; Sigala & Logothetis, 2002; van der Linden, Murre, & van Turennout, 2008). Our results add to this assertion by showing that learning-induced changes are specific to the kind of meaning that was learned.

Whereas previous fMRI research claims the RSC processes contextual information, the PHC is also strongly implicated (Aminoff et al., 2007; Bar & Aminoff, 2003; Bar et al., 2008; Ranganath & Ritchey, 2012). In experiment 1, verbal learning about the context in which an object could be found had strong effects on pattern similarity relationships in RSC, but interestingly, no comparable effects were detected in PHC. In contrast experiment 2, using real-life objects, did show contextual association effects in the PHC. This differential pattern in the PHC could be attributed to how contextual information was learned in the cases of novel and real-life objects. In experiment 1, these contextual associations were learned verbally, but never visually experienced. This contrasts with our understanding of real-world objects, for which we predominantly acquire strong contextual associations through repeated visual exposure in specific contexts. These may suggest that visual experience is key for contextual association effects in the PHC, while contextual effects in the RSC were seen for novel objects that suggests a more abstracted role of linking objects to contexts that is not as tied to visual contexts as the PHC (Bar, 2004).

### Learning-induced informational warping in the fusiform

Results from Experiment 1 demonstrated that learning meaningful information about novel objects was associated with changes in visual-form similarity relationships in the fusiform. Our anatomical ROI analysis showed that the change in the contextual association model effect over sessions was significantly greater than the change in the visual-features model effect, and critically, the reduction in the visual feature effect with learning was strongly correlated with the emergence of the contextual association effect. This result could reflect a dimensional modulation of the representational space (Folstein, Palmeri, & Gauthier, 2014) where there is a representational shift away from the pre-learning visually-based object information, to information about object meaning. This result must reflect a change in how these objects are represented in VTC, rather than a top-down effect, as the same visual task was used before and after learning that did not require specific access to object meaning. Our results further highlight how the VTC can flexibly code different kinds of object properties gained through experience (such as form and meaning) and suggests that meaning plays a critical role in shaping object representations.

### Beyond contextual association effects of meaning

Beyond learned effects of contextual associations in the fusiform, we also found evidence for more general semantic effects after learning at the spatial scale of anatomical regions. Activation patterns in the fusiform ROI correlated with the meaning and semantic category RDMs after learning which may indicate more subtle changes along semantic dimensions, where members of the same semantic category show more similar activation patterns after learning. A number of explanations could underlie the more modest semantic category effects compared to the contextual effects for novel objects. One account would be that participants in Experiment 1 learned to associate objects with labels that did not correspond to pre-existing object categories (such as animals or tools). Categories defined by their shared features (e.g. one category would be composed of metal, electrically conducting things), as in Experiment 1, have little ecological relevance, and participants could not readily leverage pre-existing category representations to facilitate learning (Op de Beeck & Baker, 2010).

This may suggest that learning about object contexts and forming superordinate categories takes place over different timescales. The significantly stronger effects for the contextual model over the semantic category model in Experiment 1 in the fusiform could be a consequence of enhanced initial learning for the contextual associations, which would be beneficial as context provides additional semantic constraints. Ad hoc testing supports the notion that contextual information aids initial learning, as our participants showed higher behavioural feature-recall accuracy for objects with a contextual ('Found in…') feature following the first learning session (session 2: contextual mean feature-recall 66%, non-contextual mean feature-recall 54%, $t(10) = 2.27$, $p = 0.046$) with this advantage disappearing with further training sessions (session 3: contextual mean feature-recall 91%, non-contextual mean feature-recall 85%, $t(10) = 2.12$, $p = 0.06$; session 4: contextual mean feature-recall 94%, non-contextual mean feature-recall 92%, $t(10) = 1.16$, $p = 0.27$). The ability to form generalisations across items that share common features could require more extensive exposure to exemplars, especially when objects are novel and cannot readily fit into existing categories. Consistent with the possibility of different timescales of context and category learning, the contextual association effect was significantly greater than the semantic category effect in the fusiform for novel objects, whereas the inverse was found for real objects. One caveat is that, in the case of real objects, semantic category and visual properties are highly confounded, whereas this was not the case for novel objects in Experiment 1. Further research is therefore needed to clarify how contextual and category information interact in the formation of stable meaningful representations in the brain.

In conclusion, the present results demonstrate that visual object representations change as a consequence of learning meaningful information for previously meaningless objects. Learning-induced changes in neural pattern similarity relationships tracked the specific kinds of information that participants learned about objects, showing how representations in the VTC can flexibly adapt to represent newly acquired meaningful information. Such changes were specific to the informational structure that was learned, further emphasising that object representations are shaped by the nature of the information we learn.

**Acknowledgements**
This project has received funding to LKT from the European Research Council (ERC) under the European Union's Horizon 2020 research and innovation programme (grant agreement No 669820), from the European Research Council under the European Community's Seventh Framework Programme (FP7/2007-2013)/ ERC Grant agreement n° 249640, and a Guggenheim Fellowship to CR.